\documentclass[preprints,article,accept,moreauthors,pdftex,10pt,a4paper]{mdpi}

\usepackage{xspace}
\usepackage{color}
\usepackage{wasysym}
\usepackage{subfigure}

\newcommand{\dd}{$\delta$D\xspace}
\newcommand{\dox}{$\delta^{18}$O\xspace}

\newcommand{\kybp}{kybp\xspace}

\firstpage{1} 
\pubvolume{xx}
\issuenum{1}
\articlenumber{5}
\pubyear{2018}
\copyrightyear{2018}
\history{Received: \today; }

\Title{ANOMALY DETECTION IN PALEOCLIMATE RECORDS USING PERMUTATION ENTROPY}

\Author{Joshua Garland$^{1,\ddagger}*$\orcidA{}, Tyler R. Jones $^{2,\ddagger}$\orcidC{}, Michael Neuder$^{3}$\orcidD{}, Valerie Morris $^{2}$\orcidF{}, James W. C. White$^{2}$\orcidE{} and Elizabeth Bradley $^{1,3}$\orcidB{} }

\AuthorNames{Joshua Garland, Tyler R. Jones, 
  Valerie Morris,  James W. C. White and Elizabeth Bradley}

\address{%
$^{1}$\quad Santa Fe Institute\\
$^{2}$\quad University of Colorado at Boulder Institute for Arctic and Alpine Research\\
$^{3}$\quad University of Colorado at Boulder Department of Computer Science\\
$^{\ddagger}$\quad These authors contributed equally to this work.} 
\corres{Correspondence: joshua@santafe.edu}

\abstract{Permutation entropy techniques can be useful in identifying
  anomalies in paleoclimate data records, including
  noise, outliers, and post-processing issues.  We demonstrate this
  using weighted and unweighted permutation entropy of
  water-isotope records in a deep polar ice core.  In one region of
  these isotope records, our previous calculations
  \cite{garland2018climate} revealed an abrupt change in the
  complexity of the traces: specifically, in the amount of \emph{new}
  information that appeared at every time step.  We conjectured that
  this effect was due to noise introduced by an older laboratory
  instrument.  In this paper, we validate that conjecture by
  re-analyzing a section of the ice core using a more-advanced version
  of the laboratory instrument.  The anomalous noise levels are absent
  from the permutation entropy traces of the new data.  In other
  sections of the core, we show that permutation entropy techniques
  can be used to identify anomalies in the raw data that are not
  associated with climatic or glaciological processes, but rather
  effects occurring during field work, laboratory analysis, or data
  post-processing.  These examples make it clear that permutation
  entropy is a useful forensic tool for identifying sections of data
  that require targeted re-analysis---and can even be useful in
  guiding that analysis.}


\begin{document}

\section{Introduction}
\label{sec:intro}
 
Paleoclimate records like ice and sediment cores provide us with long
and detailed accounts of Earth's ancient climate system. Collection of these data sets can be very expensive and the extraction
of proxy data from them is often time consuming, as well as
susceptible to both human and machine error.  Ensuring the accuracy of
these data is as challenging as it is important.  Most of these cores
are only fully measured one time and most are unique in the time period
and region that they ``observe,'' making comparisons and statistical
tests impossible.  Moreover, these records may be subject to many
different effects---known, unknown, and conjectured---between
deposition and collection.  These challenges make it very difficult to
understand how much information is actually present in these records,
how to extract it in a meaningful way, and how best to use it while not 
overusing it.

An important first step in that direction would be to identify where
the information in a proxy record appears to be missing, disturbed, or
otherwise unusual.  This knowledge could be used to flag segments of
these data sets that warrant further study: either to isolate and
repair any problems or, more excitingly, to identify hidden climate
signals.  Anomaly detection is a particularly thorny problem in
paleorecords, though.  Collecting an ice core from polar regions can
cost tens of millions of \$US.  Given the need for broad geographic
sampling in a resource-constrained environment, replicate cores from
nearby areas have been rare.  This has restricted anomaly detection
methods in paleoscience to the most simple approaches: e.g.,
discarding observations that lie beyond five standard deviations from
the mean. Laboratory technology is another issue.  Until recently, for
instance, water isotopes in ice cores could only be measured at
multi-centimeter resolution, a spacing that would lump years or even
decades worth of climate information into each data point.  The
absence of ground truth is a final challenge, particularly since the
paleoclimate evidence in the core can be obfuscated by natural
processes: material in a sediment core may be swept away by an ocean
current, for instance, and ice can be deformed by the flow of the ice
sheet.  These kinds of effects can not only destroy data, but also
deform it in ways that can create spurious signals that appear to
carry scientific meaning but are in fact meaningless to the climate record.

Thanks to new projects and advances in laboratory techniques, the resolution challenge is quickly becoming a thing of the past~\cite{cfa-las}.  In addition to the recently measured West Antarctic Ice Sheet (WAIS) Divide ice core (WDC), many additional 
high-resolution records are becoming available, such as the South Pole Ice Core (SPC)
\cite{spice} and the Eastern Greenland Ice Core Project (EGRIP) \cite{egrip}.  Replicate data is on the horizon as well, which may solve
some of the statistical challenges.  The SPC project, for instance,
will involve dual analysis (i.e. two replicate sticks of ice) from
three separate subsections of the ice core.    
However, replicate analyses of deep ice cores beyond a few hundred
meters of ice will not occur any time soon, let alone multiple cores
from a single location\footnote{There are some replicate data
  available from shallow snow pits~\cite{munch2016regional} or closely
  drilled shallow ice cores~\cite{jones2014siple}, but these do not
  extend beyond the past few hundred years, at most.}.  Thus, a
rigorous statistics-based treatment of this problem is still a distant
prospect for deep ice cores.  In the meantime, information-theoretic
methods---which can work effectively with a \emph{single} time-series
data set---can be very useful. In previous work, we showed that
estimates of the Shannon entropy rate can extract new scientific
knowledge from individual ice-core
records~\cite{IDA16,garland2018climate}.  There were hints in those
results that this family of techniques could be more generally useful in
anomaly detection.  This paper is a deeper exploration of that matter.
  
To this end, we use permutation entropy techniques to study the
water-isotope records in the WAIS Divide ice Core (WDC).  The
resolution of these data, which were measured using state-of-the-art
laboratory technology~\cite{cfa-las,WDC-isotope-data-paper}, is an
order of magnitude higher than traditionally measured deep ice core
paleoclimate records(e.g.,\cite{johnsen2001oxygen}), and also
representative of new datasets that are becoming available.  We
identify abrupt changes in the complexity of these isotope records
using sliding-window calculations of the permutation
entropy~\cite{bandt2002per} (PE) and a weighted variant of that method
known as weighted permutation entropy (WPE) that is intended to
balance noise levels and the scale of trends in the data
\cite{fadlallah2013}.  Via close examination of both the data and the
laboratory records, we map the majority of these abrupt changes to
regions of missing data and instrument error.  Guided by that mapping,
we re-measured and re-analyzed one of these segments of the core,
where the PE and WPE results suggested an increased noise level and
the laboratory records indicated that the processing had been
performed by an older version of the analysis pipeline.  The PE and
the WPE of this newly measured data---produced using state-of-the-art
equipment---are much lower, and consistent with the values in
neighboring regions of the core.  This not only validates our
conjecture that permutation entropy techniques can be used to identify
anomalies, but also suggests a general approach for improving
paleoclimate data sets in a targeted, cost-effective way.

Permutation entropy is a complexity measure: it reports the amount of
new information that appears, on the average, at each point in a
sequence.  In the context of a time series, this translates to a
measure of how information propagates forward temporally.  This has
implications for predictability~\cite{josh-pre,Pennekamp350017} among
other things; indeed, these measures have been shown to converge to
the Kolmogorov-Sinai Entropy under suitable
conditions~\cite{KELLER20121477,bandt2002entropy}, as described in
Section~\ref{sec:wpe}.  The goal here is not to measure the complexity
of paleoclimate data in any formal way, however.  Our intent, and the
underlying conjecture behind our approach, are more practical: abrupt
changes in a quantity like permutation entropy suggest that something
changed, either in the system or the data.  The following section
describes the data and methods that we use to demonstrate the efficacy
of that approach.

\section{Materials and Methods} 
\label{sec:mandm}

\subsection{Data}
\label{sec:data}

\begin{figure}[tb]
\begin{center}
\includegraphics[width=0.8\textwidth]{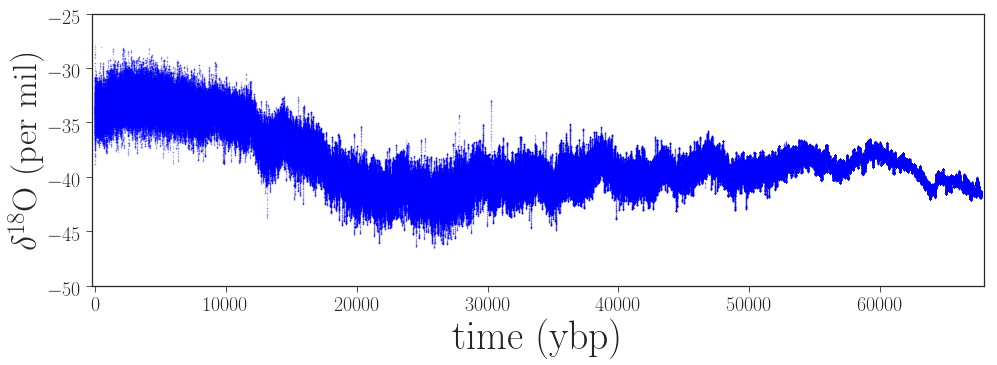}
\includegraphics[width=0.8\textwidth]{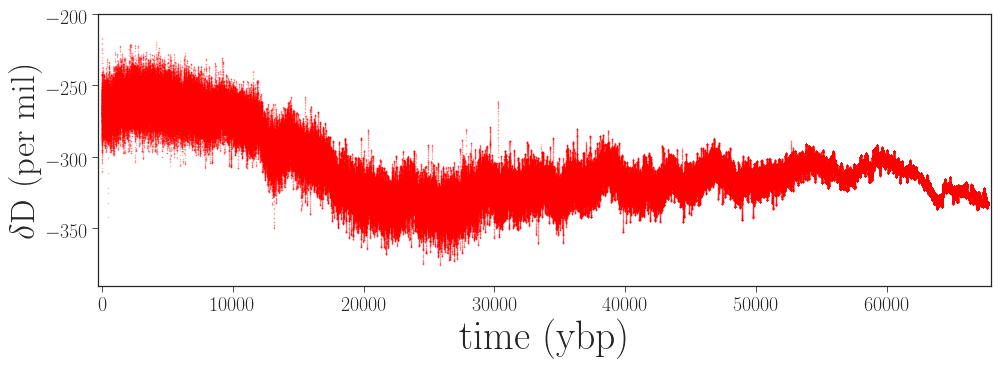}
\caption{Water-isotope records for \dox (top panel) and \dd (bottom panel) taken from the West Antarctic Ice Sheet
  (WAIS) Divide ice core (WDC).
  }
\label{fig:first}
\end{center}
\end{figure}

As a proof of concept for the claim that information theory can be useful in detecting anomalies in paleorecords, we focused on water-isotope data from the West Antarctic Ice Sheet (WAIS) Divide ice core (termed WDC hereafter)~\cite{WDC-isotope-data-paper}. These data, which consist of laser absorption spectroscopy measurements of the isotopic composition of the ice, are proxies for local temperature and regional
atmospheric circulation during the past 67,000 years.  The specific variables that we
consider are the ratios of the heavy and light isotopes of hydrogen
($^2H/^1H$) and oxygen ($^{18}$O$/^{16}$O).  Time-series traces of
these ratios, which we will identify as \dd and \dox in the rest of
this paper, are shown in Figure~\ref{fig:first}.

The details of the experiment and data processing are as follows.  The
water-isotope data were recorded at a rate of 1.18 Hz (0.85s
intervals).  Ice samples were moved through a cavity ring-down
spectroscopy continuous flow analysis (CRDS-CFA) system \cite{cfa-las}
at a rate of 2.5 cm/min, yielding millimeter resolution.  Registration
of the laser that tracks the depth along the core can be a challenge,
particularly in the brittle zone, a section of ice from approximately
$577-1300$m ($\approx 2.34-6$\kybp) in the WDC that tends to break apart
when the overburden pressure of the ice sheet is removed, allowing air
bubbles to expand and shatter the core.  This issue will return later
in this paper.  The data were then averaged over non-overlapping 0.5
cm bins. The ratio of heavy to light water isotopes in a water sample
is expressed in delta
notation~\cite{epstein1953revised,mook2000environmental} relative to
Vienna Standard Mean Ocean Water, where VSMOW has been set to
$0\permil$ by the International Atomic Energy Agency (IAEA):
$\delta=1000(R_{sample}/R_{VSMOW} - 1)$ where $R$ is the isotopic
ratio. The \dox and \dd symbols refer to fractional deviations from
VSMOW, normally expressed in parts per thousand (per mille or
\permil).

Converting the \dd and \dox data from the depth scale to an age scale
is a major undertaking, as well as a significant source of
uncertainty.  Constructing the age-depth model for the WAIS Divide
core required dozens of person-months of effort.  In the upper
$\approx 1500$ m of the core---the past 31,000 years---the age-depth
relationship for the WDC water isotope record was determined by
combining annual dating of high-resolution ($<1$cm) measurements of
electrical conductivity, di-electrical properties, and sulfur, sodium,
and black carbon concentrations\cite{Sigl16}. Below that, the
age-depth curve relies on tie points---known events, like volcanic
eruptions, that leave synchronizable signals in different climate
records---to the Hulu Cave timescale \cite{Buizert15}, with modeling
and smoothing used to fill in the gaps between those points.  This is
common practice in age models for deep ice cores, where compression
and diffusion\footnote{Water isotope records in ice cores are affected
  by diffusion~\cite{JGRF:JGRF20648}, which smooths and attenuates the
  signals.  Firn diffusion in the upper $\approx$50-100 meters of the
  ice sheet dominates these effects.
By contrast, solid diffusion is extremely slow, but over very long
time periods---tens of thousands of years---can have substantial
effects.  Mixing and diffusion also occur in the CRDS-CFA
system~\cite{cfa-las}, but these effects are small compared to firn
and solid-ice diffusion.}
eradicate the annual variations below a certain depth.  Similar issues
arise in age models for sediment cores, where marine organisms mix the
material from different time periods.  These, and other effects that
deform the timelines (e.g., brittle zone deformities introduced into
the depth scale during the registration process) are critically
important considerations when one is doing any kind of time-series
analysis on paleoclimate data sets.

The final stage in the data-processing pipeline addresses the uneven
temporal spacing of the samples.  Because of compression, 0.5 cm of
ice---the width of a sample---represents roughly 1/40th of a year of
accumulation near the top of the WDC and roughly 1.4
years near the bottom\footnote{This spacing is also affected by
  changes in yearly snow accumulation, obviously.  That effect
  actually turns out to be a major advantage, as we describe in
  \cite{garland2018climate}; together with the actions of diffusion,
  it creates a link between the accumulation rate and the information
  content that can be used to back the accumulation record out of WPE
  calculations.}.  One could certainly calculate PE or WPE on such a
sequence, but the timeline of the results would be deformed because of
the temporally irregular spacing of the data points.  Since we are
specifically interested in the time scales of the changes in
complexity, we needed to perform our calculations on data that are
evenly spaced in time, so we interpolated the \dd and \dox data to an
even 1/20th year spacing using a combination of downsampling and
linear splines.  While this is standard practice in the paleoclimate
field, it is not without problems if one plans to use
information-theoretic methods on the results, as described in more
detail in Section~\ref{sec:wpe}.  Indeed, irregular temporal spacing
is a fundamental challenge for any kind of time-series analysis
method, and the effects of interpolation methods used to regularize
the temporal spacing of these data sets must be considered very
carefully if one is doing sophisticated nonlinear statistics on the
results.  The following section describes how this plays out in the
context of WPE calculations on ice cores; several recent papers offer
more-general treatments of this matter for other kinds of paleoclimate
data
\cite{sakellariou2016counting,mccullough2016counting,npg-21-1093-2014,boers2017complete,eroglu2016see}.

\subsection{Complexity Estimation}
\label{sec:wpe}

Permutation entropy (PE)---a measure of how much new information
appears at each time step, on the average, across the computation
window---was originally proposed as a ``natural complexity measure for
a time series''~\cite{bandt2002per}.  Since its conception, PE has
been shown to converge to the Kolmogorov-Sinai
Entropy~\cite{KELLER20121477,bandt2002entropy}, or the metric Entropy
\cite{AMIGO200577,AMIGO2012789} under a variety of conditions, and
depending on features (e.g., stationary, ergodic) of the underlying
generating process.  It has also been shown to correlate with
intrinsic predictability in a variety of time
series~\cite{josh-pre,Pennekamp350017}.  For the purposes of this
paper, we simply view permutation entropy as a measure of the temporal
complexity of a time series and we treat abrupt changes in that
complexity as a signal of possible anomalies, naturally occurring or
data related.

Permutation entropy calculations combine traditional
information-theoretic probability estimates with ordinal analysis, a
process by which time-ordered elements of a time series, also known as
delay vectors $[x_t, x_{t+\tau},\dots,x_{t+(m-1)\tau}]\equiv X_{t}^{m,
  \tau}$, are mapped to an \emph{ordinal pattern}, or ``permutation,''
of the same size. If the first, second, and third points in a time series were
$X_{1}^{3,1}=[x_1,x_2,x_3]=[17.5,-1.8,4.1]$, for instance, the
corresponding permutation would be $\phi([17.5,-1.8,4.1])=231$ since
$x_2 \leq x_3 \leq x_1$. We will refer to this mapping as
$\phi:\mathbb{R}^m\to \mathcal{S}_m$, which accepts an $m$-dimensional
delay vector and returns the corresponding permutation $\pi \in
\mathcal{S}_m$, the set of permutations of length $m$. 
After using $\phi$ to convert the time-series data into a
  series of permutations, one computes the permutation entropy using
  the formula:
 \begin{equation} \label{eqn:pe}
H(m,\tau)=-\sum_{\pi\in\mathcal{S}_m}p(\pi)\log_2(\pi)
\end{equation}
\noindent where $p(\pi)$ is the estimated probability of each
permutation $\pi \in \mathcal{S}_m$ occurring in the time series, 
calculated as follows:
\begin{equation}\label{eqn:freqCount}
    p(\pi) = \frac{\left|\{t|t \leq N-(m-1)\tau, \phi(X_{t}^{m,\tau})
      = \pi\}\right|}{N-(m-1)\tau}
  \end{equation}
Here, $|\cdot|$ is set cardinality.
  
Notice that the mapping $\phi$, as defined above, does not distinguish
between $[17.5,-1.8,4.1]$ and $[17000,-1.8,4000]$; both will be mapped
to the same 231 permutation.  This may be inappropriate if the
observational noise is larger than the larger-amplitude trends in the
data, or if there is meaningful information in the amplitude of the
signal~\cite{fadlallah2013}. The now-standard way to address these concerns , \emph{weighted}
permutation entropy (WPE) \cite{fadlallah2013}, effectively emphasizes
permutations that are involved in ``large'' features and de-emphasizes
those whose amplitudes are small relative to the features of the time
series.  This is accomplished by weighting each permutation by its
variance:
\begin{equation}
  w(X_{t}^{m,\tau}) = \frac{1}{m} \sum_{j = 1}^{m}
                      \left( x_{t+(j-1)\tau} - \overline{X_{t}^{m,\tau}} \right)^2
\end{equation}
where $\overline{X_{t}^{m,\tau}}$ is the arithmetic mean of the values
in $X_{t}^{m,\tau}$.  The weighted probability of a permutation is
defined as:
\begin{equation}\label{eq:wpep}
  p_w(\pi) = \frac{\displaystyle \sum_{t \le N-(m-1)\tau} w(X_{t}^{m,\tau}) \cdot \delta(\phi(X_{t}^{m,\tau}), \pi) }{\displaystyle \sum_{t \le N-(m-1)\tau} w(X_{t}^{m,\tau})}
\end{equation}
where $\delta(x, y)$ is 1 if $x = y$ and 0 otherwise.  
The final calculation of WPE is similar to Equation~(\ref{eqn:pe})
above, but with the weighted probabilities:
\begin{equation}\label{eqn:wpe}
  H_w(m,\tau) = - \sum_{\pi \in \mathcal{S}_m} p_w(\pi) \log_2
  p_w(\pi).
\end{equation}
Following standard practice, we normalize both PE and WPE results by
dividing by $\log_2(m!)$, which causes them to range from 0 (low
complexity) to 1 (high complexity)~\cite{josh-pre}.

The temporal resolution of a permutation entropy analysis is dictated
by the span of the data across which the sum is taken.  Since we are
interested in the changing patterns in the \dd and \dox traces
described at the end of Section~\ref{sec:data}---not overall results
that are aggregated across the entire traces---the calculations in
this paper employ Equations~(\ref{eqn:pe}) and~(\ref{eqn:wpe}) in
sliding windows across those traces. 
Choosing the size $W$ for those sliding windows is not trivial.  The
minimum window width is effectively dictated by the value of $m$
because the number of data points that one needs in order to gather
representative statistics on the permutations grows with the length of
those permutations.  If, in expectation, one wants 100 chances for each
of the $m!$ possible permutations to be discovered in a given window,
there must be at least $100m!$ data points in the calculation.  But
even this minimum window size can be problematic in practice, leading
to high variances in the probability estimates from
Equations~(\ref{eqn:freqCount}) and~(\ref{eq:wpep}).  For our
purposes, this is a real issue: if $W$ is chosen too low, the high
variances can eradicate interesting features in the PE and WPE curves;
too-large $W$ values will reduce the resolution of the analysis,
thereby washing out short-time-scale anomalies.  To work around this,
it is often helpful to increase $W$ until the results stabilize, and
that is the approach taken here.
Since this will depend on the underlying statistics of the signal,
careful PE and WPE calculations require some case-by-case hand-tuning
and/or testing.  (This is true of many other data-analysis methods, of
course, though that is not widely appreciated in many scientific
fields.)

Unfortunately, the literature offers little rigorous advice regarding
how to choose $m$; the general recommendation is $3 \le m \le 6$,
without any formal justification.  The convergence proofs mentioned in
the first paragraph of this section require that $m\to\infty$; that
would require an infinitely long time series (viz., $W\geq100m!$) and
is obviously unreasonable for real-world data.  In practice, this
choice is a balance between detail and data length: short permutations
cannot capture the richness of the dynamics, but long permutations
require long calculation windows.  If $m=2$, for instance, then the
only dynamics that are captured are ``up'' and ``down.''  In a time
series with even moderate complexity, the probabilities of these two
events are usually roughly similar, so both $H(2,\tau)$ and
$H_w(2,\tau)$ saturate near 1 and neither PE nor WPE is informative.
The richer variety of permutations offered by longer word lengths can
more-accurately capture the complexity of the time series, but---as
described in the previous paragraph---that will drive up the window
size and lower the temporal resolution of the analysis.  In the face
of this, a useful practical strategy is to vary $m$ and observe the
effects on the features in the PE and WPE curves.  For real-world time
series, those features often stabilize at very low $m$ (hence the
loose recommendation cited in the first sentence of this paragraph).
In the case of the data studied here, the features of the PE and WPE
curves stabilized at $m=3$.  To be conservative, we used $m=4$ for all the
calculations reported in this paper; this, in turn, dictated a minimum
window size of 2400 points.  As a further test, we explored a range of
$m$ and $W$ values around these specific choices, confirming that the
features in the PE and WPE curves remained the same.  There is, of
course, no formal guarantee that this tuning procedure will converge
to good choices for this key free parameter, but similar
``persistence'' approaches are used in many different data-analysis
approaches (e.g., delays and dimensions for delay-coordinate
embedding~\cite{Holger-and-Liz} or length scales for topological data
analysis \cite{robins99}).

The third and final free parameter in PE and WPE calculations is the
delay, $\tau$.  In the literature, it is customary to fix $\tau=1$ and
this is precisely what we do in the anomaly-detection study reported
in this paper.  There is also serious traction to be gained by varying
the value of this parameter, though, including exploration of the time
scales on which different events occur; see \cite{garland2018climate}
for more details.

A final issue here is the timeline.  Recall that the \emph{original}
\dd and \dox measurements were spaced evenly in depth but unevenly in
time.  Because timeline deformation will obfuscate the results of
permutation entropy calculations, one must transform the measured data
to an even timeline, as described in the last paragraph of
Section~\ref{sec:data}, before invoking Equation~(\ref{eqn:pe})
or~(\ref{eqn:wpe}).  Linear interpolation, the standard practice in
paleoclimate data analysis, can be problematic in this context, as the
repeating, predictable patterns introduced by interpolation can skew
the distribution of the permutations and thereby lower the PE and WPE
values\footnote{One could also downsample the original measurements to
  a fixed temporal spacing, but that would involve discarding large
  amounts of the data in the top layers of the core, and thus would
  greatly reduce the resolution of the resulting WPE analysis.}.  This
effect will generally worsen with depth because more interpolation is
required lower in the core, where the temporal spacing between the
measured samples is larger.  As mentioned in Section~\ref{sec:data},
the uncertainty of the age scale also comes into play here---another
effect that worsens with depth~\cite{npg-21-1093-2014}.  For all of
these reasons, one cannot compare WPE values across wide ranges of the
timeline of an ice core.
 
Permutation entropy techniques have been used successfully to detect a
variety of changes in time series: e.g., epileptic seizures in EEG
signals~\cite{cao2004det}, bifurcations in the transient logistic
map~\cite{cao2004det}, voiced sounds in a noisy speech signal
\cite{bandt2002per}, and market inefficiencies in financial records
\cite{ZUNINO20092854,ZUNINO20101891,forbidden-finance}.  They have
also been used to assess predictability of time series in a variety of
fields \cite{josh-pre,Pennekamp350017} and reveal various interesting
effects in paleoclimate
data~\cite{IDA16,garland2018climate,Saco2010,balasis2013statistical}.
The bulk of this work has used the \emph{weighted} variant of the
technique.  As we will show in this paper, though, PE and WPE work
together in a complementary fashion, detecting different kinds of
anomalies in paleoclimate records.

\section{Results}
\label{sec:results}

\begin{figure}[t]
\begin{center}
    \begin{subfigure}
        \centering
        \includegraphics[width=0.8\textwidth]{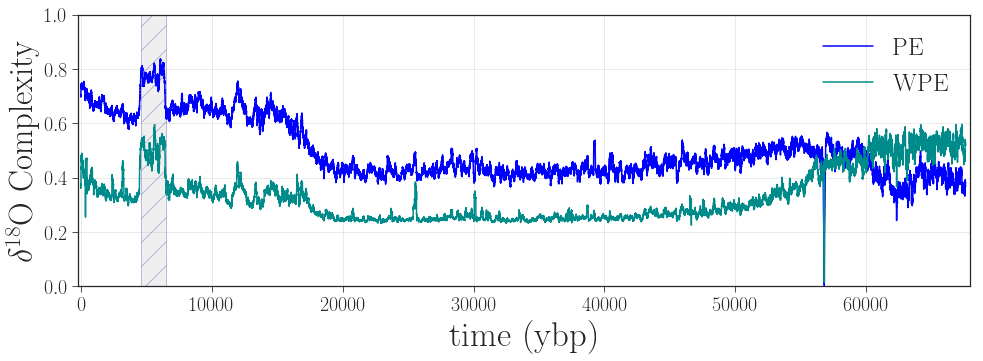}
    \end{subfigure}%
    \begin{subfigure}
        \centering
    \includegraphics[width=0.8\textwidth]{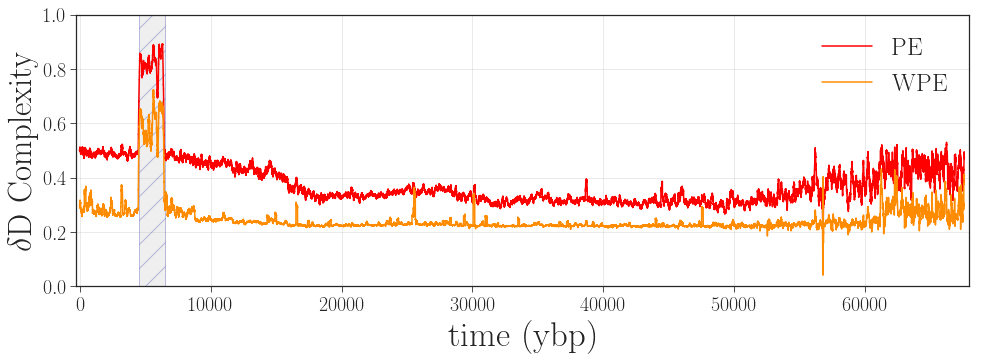}
    \end{subfigure}%
\caption{Permutation entropy of the \dox and \dd data from
  Figure~\ref{fig:first}, calculated in rolling 2400-point windows
  (i.e., $W=2400$) with a word length $m=4$ and a delay $\tau=1$.  Top
  panel: weighted and unweighted permutation entropy of \dox in cyan
  and blue, respectively.  Bottom panel: WPE and PE of \dd in orange
  and red, respectively.  }
\label{fig:originalWPE}
\end{center}
\end{figure}

Figure~\ref{fig:originalWPE} shows the weighted and unweighted
permutation entropies of the \dox and \dd data from
Figure~\ref{fig:first}.
There are a number of interesting features in these four curves, many
of which have scientific implications, as discussed
in~\cite{IDA16,garland2018climate}.  Here, we will focus on apparent
anomalies in the complexity of these climate signals: i.e.,
discontinuities or abrupt changes in the curves.  The most obvious
such feature is the large jump in all four traces between
$\approx$4.5-6.5 \kybp, shown shaded in grey in
Figure~\ref{fig:originalWPE}.  This sharp increase in the complexity
of both signals indicates that something is fundamentally different
about this segment of the record, compared to the surrounding regions:
in particular, that the values of both \dd and \dox depend less on
their previous values during this period than in surrounding
regions---i.e., that a large amount of new information is produced at
each time step by the system that generated the data.
There are two potential culprits for such an increase: data issues or
a pair of radical, rapid climate shifts at either end of this time
period.  Since no such shifts are known or hypothesized by the
climate-science community, we conjectured that this jump was due to
processing-related issues in the data.

Recall that PE and WPE are designed to bring out different aspects of
the information in the data.  In view of this, the fact that
\emph{both} calculations pick up on this particular feature is
meaningful: it suggests that the underlying issue is not just
noise---which would be at least partially washed out by the weighting
term, causing the jumps in WPE to be smaller than those in PE.
Rather, there may be other causes at work here.
Going back to the laboratory records and the original papers, we found
that an older, less precise, instrument was used to analyze this
section of ice, and that this region of the core received the poorest
possible quality score~\cite{Sigl16,souney2014core}.  This part of the
core is from the aforementioned brittle zone, which shatters when
removed from the ice sheet.  This not only makes the timeline
difficult to establish, as mentioned in Section~\ref{sec:data}, but
can also affect the \dd and \dox values, as the drilling fluid can
penetrate the ice and contaminate the measurements~\cite{Sigl16}.
This could easily disturb the amplitude-encoded information in the
core.  Unfortunately, permutation entropy techniques cannot tell us
what the underlying causes of this anomaly are; that requires expert
analysis and laboratory records.  Even so, their ability to flag
problems, and help experts form scientific hypotheses about their
causes, is a major advantage of these information-theoretic techniques.

\begin{figure}[tb]
\begin{center}
\includegraphics[width=0.45\textwidth]{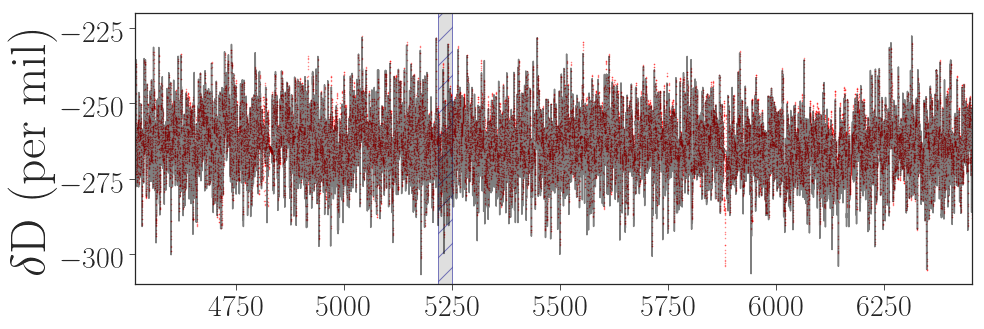}
\includegraphics[width=0.45\textwidth]{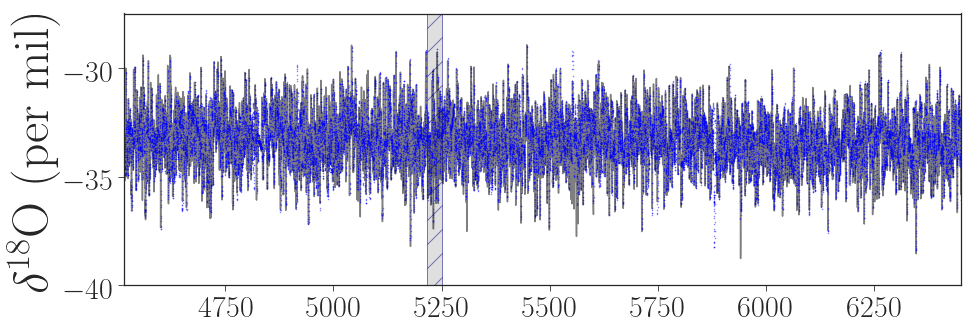}
\includegraphics[width=0.45\textwidth]{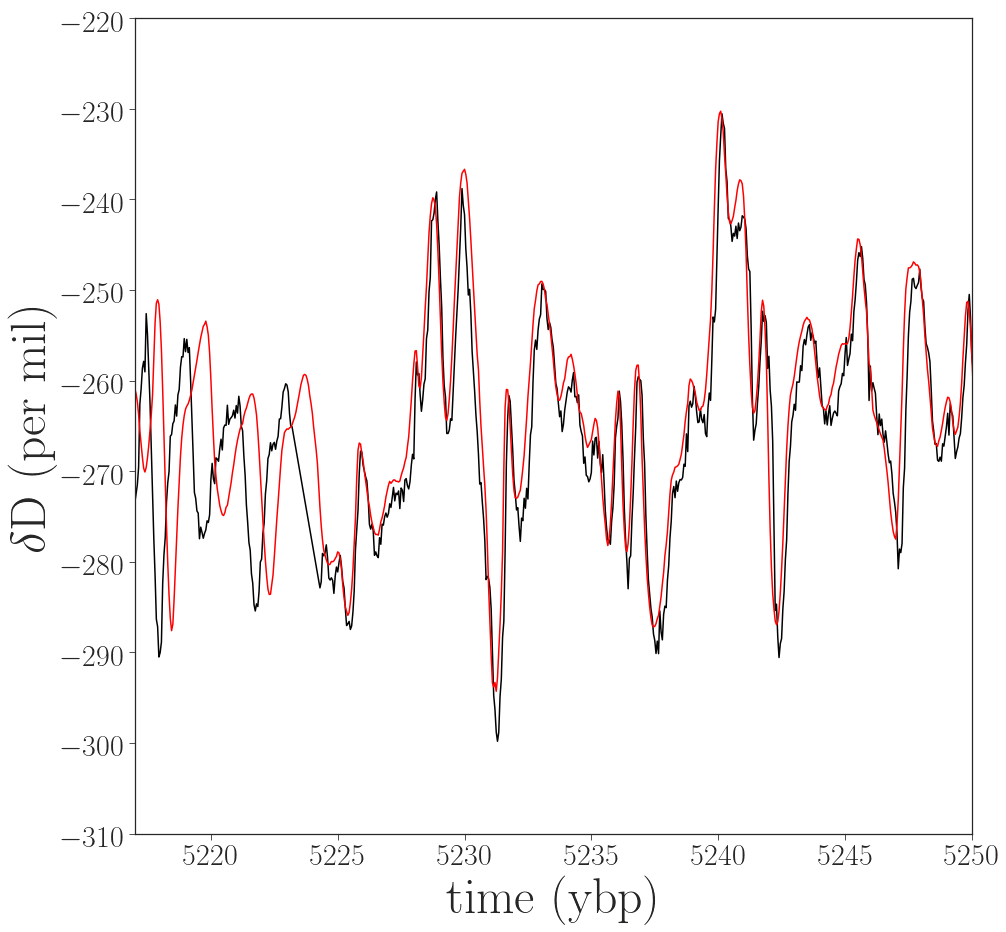}
\includegraphics[width=0.45\textwidth]{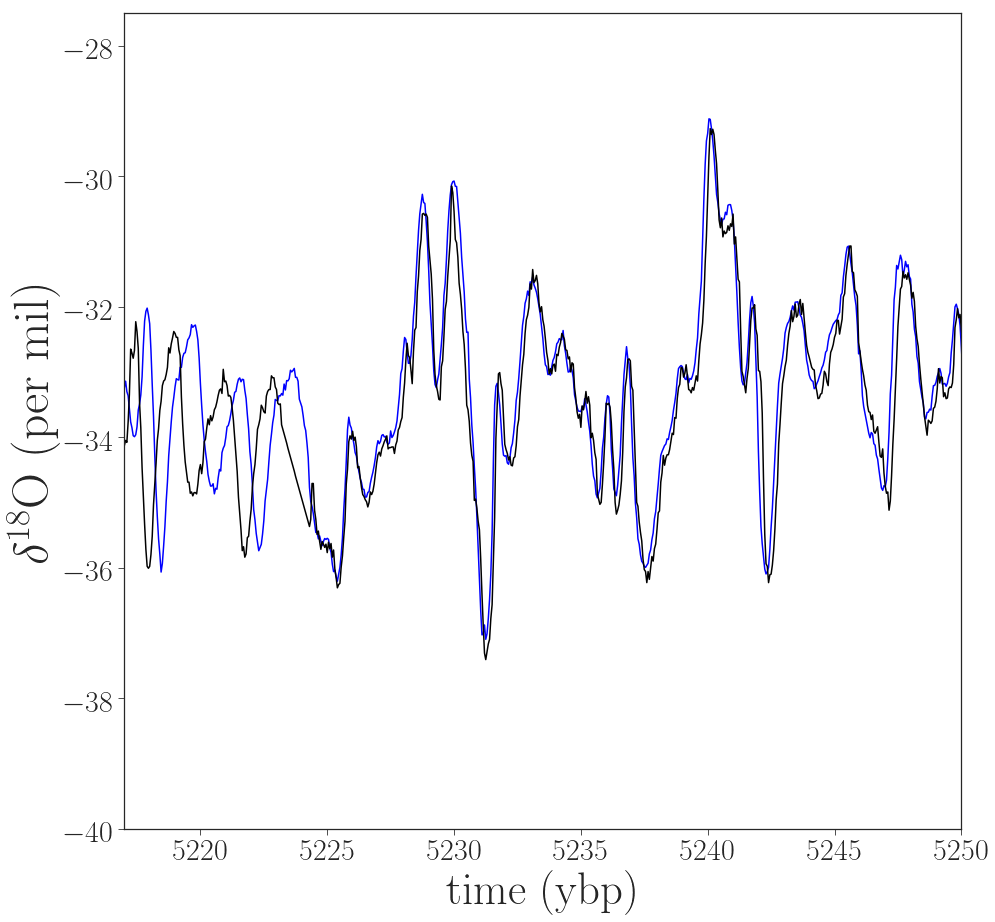}
\caption{Top panels: the re-measured \dd (red) and \dox (blue) data
  points in the range $\approx$4.5-6.5 \kybp, with the original
  records shown in grey.  Bottom panels: a closeup of a small segment
  of the traces, marked with a grey box in the top panels. Note that the vertical scales here are different than in
  Figure~\ref{fig:first}.
}
\label{fig:replace}
\end{center}
\end{figure}

As a case in point, we tested our hypotheses about the grey-shaded
jump in Figure~\ref{fig:originalWPE}
by re-measuring the 331 m segment of the WAIS Divide ice core that
corresponds to this time period.  After obtaining the archived ice at
the National Science Foundation Ice Core Facility (NSF-ICF), we
re-measured the isotope data with state-of-the-art equipment,
repeating the depth-to-age conversion and temporal regularization
processes described in Section~\ref{sec:mandm}.  Plots of these new
traces appear in Figure~\ref{fig:replace}, with the old values shown
in black.
Visual examination of these traces makes two things apparent: lower
levels of high-frequency noise in the new signals
and small phase offsets between the old and new data.  The lower noise
levels were, of course, the point of the resampling.  The minor phase
disparities are a consequence of updates in the laboratory pipeline
and the difficulties posed by this particular section of the core.  In
the years since the creation of the first generation of this
system---which was used to measure the original data---there have been
a number of updates and improvements.  As a result, there are subtle
differences (on the order of a few centimeters) in the depth
registration between the re-measured and the original data.  This is a
particular challenge in the brittle zone, since the broken or
shattered ice pieces are difficult to piece back together and can
settle along fractures, reducing the length of the ice stick by a few
centimeters.
This can even cause complete data loss in short segments.
Re-measuring this ice using state-of-the-art technology has allowed us
to improve the data in several important ways, by reducing
the overall noise levels, improving the depth registration,
and even filling in missing parts of the original record that we
obtained obtained upon remeasuring the archived ice.

\begin{figure}[tb]
\begin{center}
\includegraphics[width=0.8\textwidth]{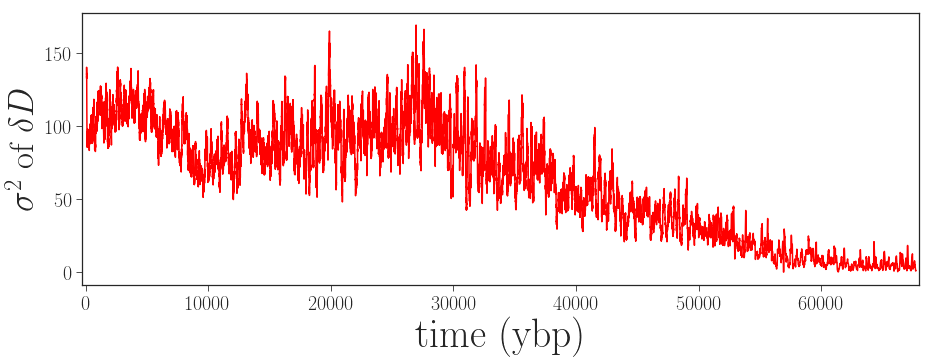}
\includegraphics[width=0.8\textwidth]{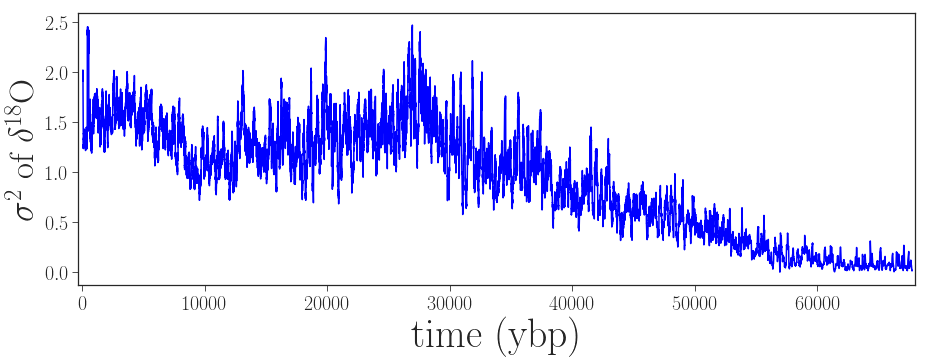}
\caption{A simple anomaly detection algorithm.  Here $\sigma^2$ is
  estimated on a rolling 2400-point overlapping window for \dd (top)
  and \dox (bottom).  Neither the feature in the grey-shaded box in
  Figure~\ref{fig:originalWPE}, nor the other anomalies described
  later in this paper, are brought out by this calculation. }
\label{fig:var}
\end{center}
\end{figure}

Given the kind of issues that were present in the data, such as
increased small-scale variance, it is worth considering whether
simpler approaches---less computationally complex than permutation
entropy, and with fewer free parameters---would be equally effective
in flagging the grey-shaded jump in Figure~\ref{fig:originalWPE}.
Because of the inherent data challenges (unknown processes at work on
the data, lack of replicates, laboratory issues, etc.), this community
has been traditionally limited to fairly rudimentary approaches to
anomaly detection: e.g., discarding observations that lie beyond five
standard deviations from the mean.  Accordingly, we performed a
rolling-variance calculation on the same isotope records using a
2400-point window.  These results, shown in Figure~\ref{fig:var}, do
not bring out the $\approx$4.5-6.5 \kybp feature, nor do they identify
the other anomalies that are described later in this paper.
Of course, there are surely other anomaly-detection methods that could
be useful in identifying anomalies in paleoclimate records.  However,
permutation entropy methods identify these problems quite clearly, are
relatively simple and fast to compute and, while they have several
parameters to tune, the results are quite robust to these choices.

\begin{figure}[t]
\begin{center}
    \begin{subfigure}
        \centering
        \includegraphics[width=0.8\textwidth]{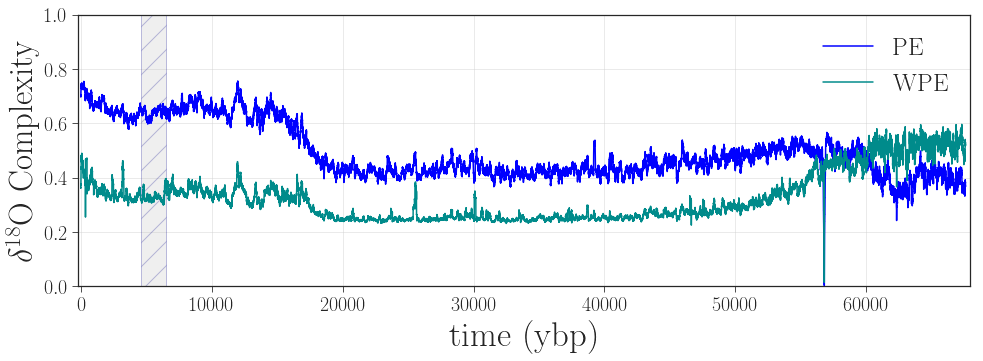}
    \end{subfigure}%
    \begin{subfigure}
        \centering
    \includegraphics[width=0.8\textwidth]{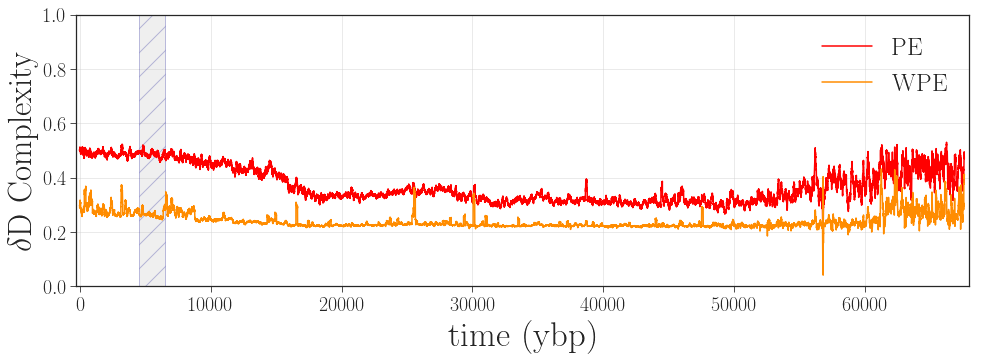}
    \end{subfigure}%
\caption{PE and WPE of \dox (top panel) and \dd (bottom panel) using
  re-measured data from 1037-1368m ($\approx4.5-6.5$ \kybp).}
\label{fig:remeasuredWPE}
\end{center}
\end{figure}

The last step in validating that the jump in the curves in
Figure~\ref{fig:originalWPE} indicated an anomaly
was to replace the data points between 4.5-6.5 \kybp in the
\dd and \dox records of Figure~\ref{fig:first} with the re-measured
values from Figure~\ref{fig:replace} and repeat the PE and WPE
calculations.  The results are quite striking, as shown in
Figure~\ref{fig:remeasuredWPE}: the large square waves are completely
absent from the PE and WPE traces of the repaired data set.
This experiment---the longest high-resolution re-measurement and re-analysis of an ice core that has
been performed to date---not only confirms our conjecture that this
anomaly was due to instrument error, but also allowed us to improve an
important section of the data using a targeted, highly focused effort.

 The repaired segment of water isotope data ($\approx4.5-6.5$ \kybp) captures climate signal from the Holocene era, 
and thus may contain useful information about the onset of climactic
shifts during the beginnings of human civilization.  In view of this,
recognizing and repairing issues with these data is particularly
meaningful.  Again, permutation entropy calculations alone cannot
tell us the \emph{underlying mechanism} for abrupt changes.  However,
their ability to identify a region of the core that should be
revisited---because of issues that would only have been apparent with
a laborious, fine-grained traditional analysis of the data---is a
major advantage.  This result highlights the main point that we wish
to make in this paper: PE and WPE are useful methods for identifying
anomalies in time-series data from paleoclimate records.  This is
particularly useful in contexts where the data are difficult and/or
expensive to collect and analyze.  In these situations, a tightly
focused new study, specifically targeted on the offending area using
permutation entropy, can maximize benefit with minimum effort.

Another obvious feature in Figure~\ref{fig:remeasuredWPE} is the
downward spike around 58 \kybp in all four curves.  Alerted by this
anomaly in WPE and PE, we returned to the laboratory records and found
that 1.107 m (110.1 yr) of ice was unavailable from the record in this
region and so a span of $\approx$ 2387 points in the \dd and \dox
traces were filled in by interpolation.  This series of points---a
linear ramp with positive slope---translated to a long series of
``1234'' permutations.  This causes a drop in the PE curves as the
calculation window passes across this expanse of completely
predictable values.  Indeed, for calculations with $\tau = 1$ and $W =
2400$, there is a brief period where 99.45\% of the "data'' in that
window has the same permutation, which causes WPE to fall
precipitously, then spike back up as the window starts to move back
onto non-interpolated data.  

\begin{figure}[tb]
\begin{center}
\includegraphics[width=\textwidth]{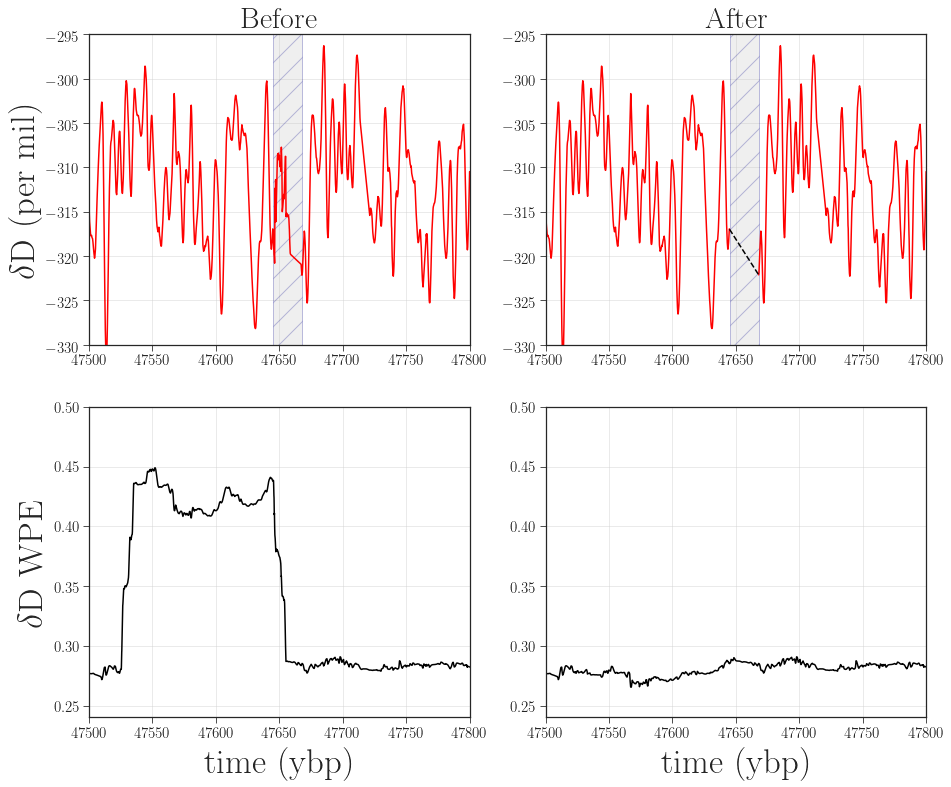}
\caption{Top row: \dd data from 47.5-47.8 \kybp before and after
  removal of and interpolation over (black dashed line) a range of
  faulty values (shaded in gray).  Bottom row: WPE calculated from the
  corresponding traces.  The width of the square wave in the lower
  left plot is the size of the WPE calculation window (2400 points at
  1/20th year per point) plus the width of the anomaly.  The
  horizontal shift between the earliest faulty value and the rise in
  WPE is due to the windowed nature of the WPE calculation.}
\label{fig:h-bumps-wpe}
\end{center}
\end{figure}

The large jump from $\approx4.5-6.5$ \kybp and the spike at 58 \kybp
are only a few of the abrupt changes in the permutation entropy
traces.  The other spikes and dips in those curves, we believe, are
also associated with anomalies in the data.  Some of these features
appear in both PE and WPE; others appear in PE but not WPE (e.g., at
38.7 \kybp in \dd) or vice versa (e.g., 25.6, 30.1, and 47.5 \kybp for
\dd and \dox).  This brings out the differential nature of these two
techniques, and the power of the combined analysis: each can,
independently, detect types of anomalies that the other misses.  In
the remainder of this section, we dissect a representative subset of
the anomalies in Figure~\ref{fig:remeasuredWPE} to illustrate this
claim, beginning with the case where WPE, but not PE, suggests data
issues.
Manual re-examination of the data in the regions around 25.6, 30.1,
and 47.5 \kybp revealed that the \dd and \dox signal in these regions
had been compromised in very small segments of the record ($<1m$ of
ice).  The top left panel of Figure~\ref{fig:h-bumps-wpe} shows the
region around 47.5 \kybp.  The signal is visibly different in the
small shaded region, combining sharp corners and high-frequency
oscillations---unlike the comparatively smooth signal in the regions
before and after this segment.
WPE is much better at picking up this kind of anomaly because there
has been a drastic shift in the information that is encoded in the
\emph{amplitude} of the signal.  PE, in contrast, does not flag this
segment because the distribution of permutations in this region is
similar to that of original signal. However, the variance of each
delay vector is quite different than in the surrounding signal: an
effect to which PE is blind, but that WPE is designed to emphasize.

There are many potential mechanisms that could be causing this
distortion in the amplitude: extreme events, for instance, such as
large volcanic eruptions, or the same kinds of data and
data-processing issues that are discussed above.  Returning again to
the laboratory records, we discovered that the CRDS-CFA system had
malfunctioned while analyzing this region.  We could, as before,
confirm that this was the cause of the anomaly by re-measuring this
region of the core.  If the outliers disappeared as a result, that
would resolve some significant data issues with only a minimal,
targeted amount of effort and expense.  If they did \emph{not}
disappear, then the permutation entropy calculations would have
identified a region where the record warranted further investigation
by paleoclimate experts.  To date, we have not carried out this
experiment, but plan to request archived ice from the NSF-ICF for
re-measurement and re-analysis.
Access to this limited and irreplaceable resource is, justifiably so,
highly guarded---particularly in regard to the deeper ice.  In lieu of
that experiment, we re-processed the data from the region surrounding
$47.5-47.8$ \kybp in the \dd record by removing the offending data
points and interpolating across the interval.  As shown in the
bottom-right panel of Figure~\ref{fig:h-bumps-wpe}, this removes the
small square wave from the WPE trace.  Without being able to
re-measure this ice, of course, we cannot narrow down the cause of
this anomaly.  Even so, the WPE results are useful in that they allow
us to do some targeted reprocessing of the data in order to mitigate
the effects.

\begin{figure}[tb]
\begin{center}
\includegraphics[width=0.8\textwidth]{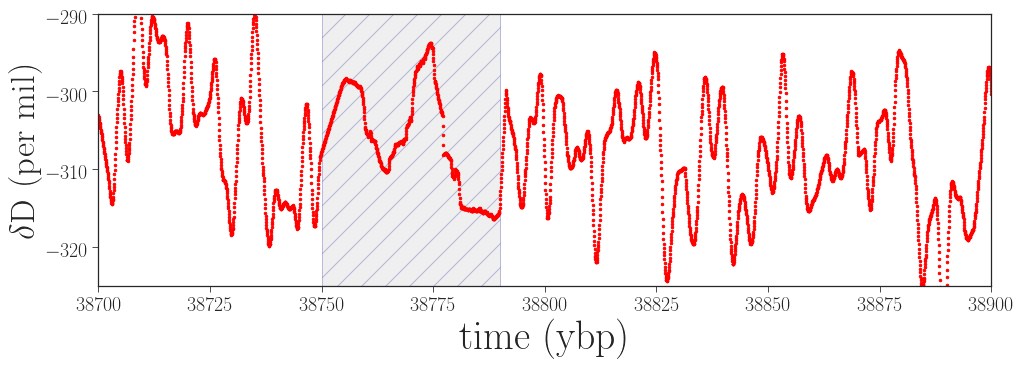}
\includegraphics[width=0.8\textwidth]{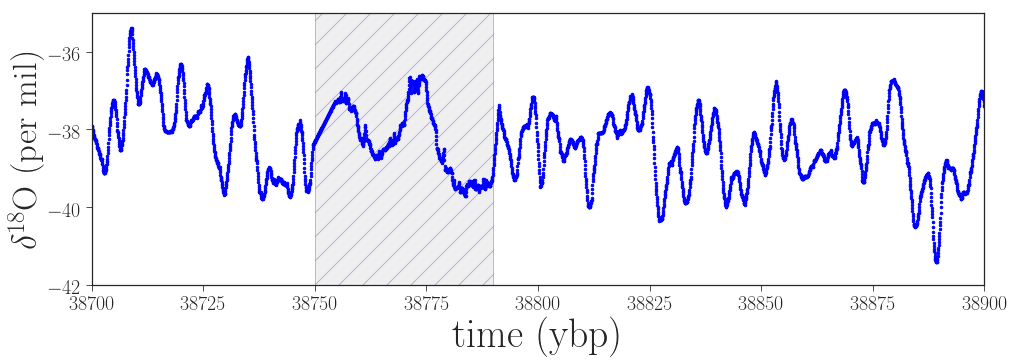}
\caption{Isotope data near 38.7 \kybp that produced a spike in PE
  but not in WPE.  According to the lab records, the graphical user
  interface froze during the analysis of this segment of the ice,
  compromising the results.}
\label{fig:h-bumps-pe}
\end{center}
\end{figure}

Spikes in PE that are \emph{not} associated with equally strong spikes
in WPE are indications of effects that are dominated by small-scale
noise.  Figure~\ref{fig:h-bumps-pe} shows an example of the isotope
record in one of these regions.
In contrast to the situation in Figure~\ref{fig:h-bumps-wpe},
which is dominated by high-frequency oscillations, this anomaly
takes the form of a strong low-frequency component with a small
band of superimposed noise.  The weighting strategy in WPE causes
it to ignore these features, but PE picks up on them very
clearly\footnote{Indeed, the frequency shift may magnify the
impact of noise on that calculation.}.  To diagnose the cause of
this anomaly, we again returned to the lab records, finding that
the graphical user interface froze while analyzing this
particular ice stick, and that the data-processing routines
involved in distance registration and isotope content were
compromised.

Finally, there is a single spike (near 16.6 \kybp in the WPE of \dd)
where visual inspection of the raw data by expert paleoscientists does
not suggest any issues, nor are there any recorded problems in the
laboratory records.  This is not part of any sections of the core that
are known to be problematic, like the brittle zone, and most likely
warrants further investigation.

\section{Discussion}
\label{sec:discussion}

This manuscript illustrates the potential of information theory as a
forensic tool for paleoclimate data, identifying regions with
anomalous amounts of complexity so that they can be investigated
further.  As a proof-of-concept demonstration of this claim, we used
permutation entropy techniques to isolate issues in water-isotope data
from the WAIS Divide core, then re-measured and re-analyzed one of the
associated regions using state-of-the-art laboratory equipment.  The
results of this experiment---the longest segment of replicate ice-core
analysis in history---verified that the issue flagged by these
information-theoretic techniques in that region were caused by
outdated laboratory techniques, as we had conjectured.  We also showed
that permutation entropy techniques identified several other smaller
anomalies throughout the record, which we showed were caused by
various minor mishaps in the data processing pipeline. Finally, unlike
other anomaly-detection studies that use either PE or WPE, we showed
that using these two techniques together, rather than in isolation,
allows detection of a much richer landscape of possible anomalies.

This re-analysis approach is already providing valuable insights that
reach well beyond the scope of this work.  Compared to the prohibitive
cost and time commitment of collecting a new core, re-measuring ice
from an archived core can be comparatively easy and inexpensive if the
region requiring re-analysis can be precisely and clearly defined.
This forms the basis for what we believe will be an effective
\emph{general} targeted re-investigation methodology that can be
applied not only to the water-isotope record from the WAIS Divide ice
core, but to other high-resolution paleoclimate records.  Indeed,
several such records are being finalized at the time of this
publication, which makes this a perfect time to start developing,
applying, and evaluating sophisticated anomaly-detection methods for
these important data sets.  Excitingly, as a result of the study
reported in this manuscript, PE-based techniques are now being added
to the quality control phase of the analysis pipelin for at least one
of these new high-resolution records.

The information in paleoclimate records contains valuable clues and
insights about the Earth's past climate, and perhaps about its future.
While these records have a lot of promise, it is crucial---especially
in an era of rampant climate-change denial---to be careful in
extracting the useful and meaningful information from these records
while simultaneously identifying regions that are problematic.  For a
multitude of reasons, distinguishing useful information from a lack
thereof can be a particularly challenging task in paleodata.  This is
especially true in parts of these records that are hard to analyze,
such as the brittle zone of the WDC.  Our work has identified a number
of intervals in this core where the data require a closer look,
including several that contain information corresponding to the time
period of the dawn of human civilization.

\vspace{6pt}

\authorcontributions{conceptualization, J.G.; methodology, J.G., T.J., E.B., M.N. and J.W.; software, J.G. and M.N..; validation, J.G., T.J., E.B., M.N. and J.W.; formal analysis, J.G., E.B. and T.J.; investigation, J.G., T.J, M.N.,E.B. and V.M.; resources, T.J, J.W. and V.M.; data curation, V.M., J.W. and T.J.;writing---original draft preparation J.G. and E.B.; writing---review and editing, J.G., E.B. and T.J.; visualization, J.G. and M.N.; supervision, E.B. and J.W..; project administration, J.G. and E.B.; funding acquisition, J.G. and T.J.}

\funding{This research was funded by US National Science Foundation
  (NSF) grant number 1807478. JG was also supported by an Omidyar Fellowship at the Santa Fe Institute. For original ice core data, this work
  was supported by NSF grants 0537593, 0537661, 0537930, 0539232,
  1043092, 1043167, 1043518 and 1142166.  Field and logistical
  activities were managed by the WAIS Divide Science Coordination
  Office at the Desert Research Institute, USA, and the University of
  New Hampshire, USA (NSF grants 0230396, 0440817, 0944266 and
  0944348). The NSF Division of Polar Programs funded the Ice Drilling
  Program Office (IDPO), the Ice Drilling Design and Operations (IDDO)
  group, the National Science Foundation Ice Core Facility (NSF-ICF), the Antarctic
  Support Contractor, and the 109th New York Air National Guard.}

\acknowledgments{Water isotope measurements were performed at the
  Stable Isotope Lab (SIL) at the Institute of Arctic and Alpine
  Research (INSTAAR), University of Colorado. We wish to thank Bruce
  Vaughn for his efforts in designing and implementing the
  laser-based, continuous flow measurement system for water isotopes,
  as well as Mark Twickler, Geoff Hargreaves, and Richard Nunn for
  their assistance planning and executing ice sampling at NSF-ICF.  We
  thank Bedartha Goswami for his advice. We would also like to thank
  the reviewers of this paper for their valuable
  feedback---particularly Peter Ditlevsen. }

\conflictsofinterest{The authors declare no conflict of interest.} 

\abbreviations{The following abbreviations are used in this manuscript:\\

\noindent 
\begin{tabular}{@{}ll}
CFA & Continuous Flow Analysis \\
CRDS-CFA & Cavity Ring-Down Spectroscopy Continuous Flow Analysis \\
NSF-ICF & National Science Foundation Ice Core Facility \\
PE & Permutation Entropy\\
WPE & Weighted Permutation Entropy\\
WAIS & West Antarctic Ice Sheet\\
WDC & WAIS Divide Ice Core
\end{tabular}}


\reftitle{References}


\bibliography{master-refs}

\end{document}